\documentclass[%
 reprint,
superscriptaddress,
 amsmath,amssymb,
 prx,
]{revtex4-2}

\usepackage{graphicx}
\usepackage{dcolumn}
\usepackage{bbm}
\usepackage{braket}
\usepackage{xcolor}
\newcommand{\eph}{e_{\rm{ph}}}

\usepackage{graphicx}
\usepackage{dcolumn}
\usepackage{bm}

\begin{document}

\preprint{APS/123-QED}

\title{Securing practical quantum communication systems with optical power limiters}

\author{Gong Zhang}
 \email{zhanggong@nus.edu.sg}
\affiliation{Department of Electrical \& Computer Engineering, National University of Singapore, Singapore}

\author{Ignatius William Primaatmaja}
\affiliation{Centre for Quantum Technologies, National University of Singapore, Singapore}
\author{Jing Yan Haw}
\affiliation{Department of Electrical \& Computer Engineering, National University of Singapore, Singapore}
\author{\\Xiao Gong}
\author{Chao Wang}
 \email{wang.chao@nus.edu.sg}
\affiliation{Department of Electrical \& Computer Engineering, National University of Singapore, Singapore}

\author{Charles Ci Wen Lim}
 \email{charles.lim@nus.edu.sg}
\affiliation{Department of Electrical \& Computer Engineering, National University of Singapore, Singapore}
\affiliation{Centre for Quantum Technologies, National University of Singapore, Singapore}

\date{\today}

\begin{abstract}
Controlling the energy of unauthorized light signals in a quantum cryptosystem is an essential criterion for implementation security.
Here, we propose a passive optical power limiter device based on thermo-optical defocusing effects providing a reliable power limiting threshold which can be readily adjusted to suit various quantum applications. In addition, the device is robust against a wide variety of signal variations (e.g. wavelength, pulse width), which is important for implementation security. 
Moreover, we experimentally show that the proposed device does not compromise quantum communication signals, in that it has only a very minimal impact (if not, negligible impact) on the intensity, phase, or polarization degrees of freedom of the photon, thus making it suitable for general communication purposes. To show its practical utility for quantum cryptography, we demonstrate and discuss three potential applications: (1) measurement-device-independent quantum key distribution with enhanced security against a general class of Trojan-horse attacks, (2) using the power limiter as a countermeasure against bright illumination attacks, and (3) the application of power limiters to potentially enhance the implementation security of plug-and-play quantum key distribution. 
\end{abstract}

\maketitle

\section{Introduction}
Quantum key distribution (QKD) enables two remote network users to exchange provably-secure keys when it is implemented faithfully~\cite{lo2014secure,Diamanti2016,XuRMP2020}. To ensure implementation security, the research community has been focusing on the security of practical systems in recent years,
developing methods to narrow the gap between the theory and practice of QKD. 
On the theoretical side, robust QKD protocols have been proposed, which not only make practical systems more secure against device imperfections but also easier to calibrate and validate in practice (since fewer assumptions are required). 
On the experimental side, efforts have been focused on tackling quantum side-channels and a wide variety of countermeasures have been proposed and developed~\cite{Nitin2016,XuRMP2020}. 

Trojan-horse attacks (THAs)~\cite{Vakhitov2001, gisin_trojan-horse_2006} represent one of the biggest threats to QKD security. These attacks aim to steal the secret key information via the injection of unauthorized light pulses, seeking to carry critical modulation information out of the transmitters. More specifically, in these attacks, the adversary (henceforth called Eve) injects bright light pulses into the transmitter and collects the reflected light pulses. Consequently, this allows Eve to learn some information about the secret key. It has been shown that these kind of attacks can be readily implemented using standard optical methods~\cite{Vakhitov2001, gisin_trojan-horse_2006,jain2014trojan, Sajeed2017}. To mitigate this issue, one can use specialized security analyses to include security against specific types of THAs; for instance, by modeling the unauthorized input light pulses as coherent states. Then, under the assumption that the energy of the reflected light pulses is bounded, one can compute the secret key rate, as was done in~Refs.~\cite{lucamarini_practical_2015,tamaki_decoy-state_2016}.

The bright illumination attacks are another particularly powerful class of side-channel attacks. These include laser damage attacks~\cite{bugge2014laser,makarov_creation_2016,huang_laser-damage_2020} and blinding attacks~\cite{makarov2009controlling,lydersen2010hacking,yuan_resilience_2011}. In these attacks, bright light pulses are used to control QKD devices by exploiting their implementation knowledge. Consequently, these allow Eve to avoid eavesdropping detection and hence security is no longer guaranteed. Fortunately, there exist countermeasures which are pretty effective against such attacks~\cite{yuan_avoiding_2010,yuan_resilience_2011} and innovative QKD protocols which are completely immune against detection side-channel attacks are known as well, e.g., see measurement-device-independent QKD (MDI QKD)~\cite{Lo2012, Braunstein2012}.

Based on the above, it can therefore be said that the injection of (unauthorized) bright light pulses into quantum communication systems is a catalyst for side-channel attacks. This is not so surprising since the presence of bright light pulses essentially breaks one of the most important assumptions of quantum cryptography---that the energy of the underlying quantum signals is at the single-photon level (or sufficiently small). To overcome these potential loopholes, one promising solution is to limit the energy of incoming light. Indeed, if this is achieved, one can be sure that the QKD system is operating at the single-photon level and the energy of any outgoing light pulse is bounded as well. Consequently, this will allow the system to operate faithfully in the quantum regime. 

In practice, this solution would mean introducing a kind of \emph{quantum power limiting} device into the QKD system. Based on current research on side-channel attacks, we believe an ideal quantum power limiter should possess the following properties: (a) able to provide a reliable and adjustable photon energy limiting down to the order of a few photons to hundreds of photons for each quantum state, (b) have a minimum insertion loss if the input power is below the threshold and stop the transmission or maintain at the threshold power once the input power exceeds the threshold, and (c) the power limiting effects are independent of other physical degrees of freedoms, e.g., frequency, polarization, etc. In terms of practical considerations, the power limiter device should also be cost-effective, passive, and easily replaceable (if it cannot recover to its normal state after being exposed to strong light).

Here, we propose and demonstrate a novel and practical quantum power limiter that can secure a broad class of QKD setups~\cite{patent}. The device is based on a form of thermo-optical defocusing effect, which effectively bounds the output optical power by some predetermined threshold. By modeling the system using a set of physically relevant assumptions, we show that the output-input optical power relation of the proposed device can be precisely controlled by changing the system parameters, e.g. the length of the prism and the diaphragm width. Consequently, this allows us to tailor the device to different quantum cryptographic applications. The feasibility and performance of our proposed power limiter device are confirmed using COMSOL (a multi-physics simulation software) and experimental data.

The paper is organized as follows. In Section.~\ref{sect2}, we first present the design details and the modeling of our power limiter. Thereafter, simulation and experimental results are illustrated. Section.~\ref{sect3} discusses the potential implementation loopholes and the robustness of the proposed power limiter. 
In Section.~\ref{sect3.5}, we experimentally verify that the power limiter is essentially transparent to standard quantum encoding choices such as intensity, phase, and polarization degrees of freedom.
In Section.~\ref{sect4}, we illustrate the broad utility of the proposed power limiter over three different QKD systems. In the first application, we provide a general security analysis of MDI QKD that allows for Eve to inject in any kind of state in a given Trojan-horse optical mode. Thereafter, a detailed study on the application of our power limiter in MDI QKD is presented, followed by the simulation results. In the second and third applications, we discuss how the proposed power limiter could be utilized to deter bright illumination attacks and to enhance the implementation security of plug-and-play QKD~\cite{muller_plug_1997, xu_measurement-device-independent_2015, tang2016experimental}. In Section.~\ref{sect5} we end with a conclusion.

\section{Optical power limiter design}\label{sect2}

\begin{figure}[tbp]
\centering
\includegraphics[width=1\columnwidth]{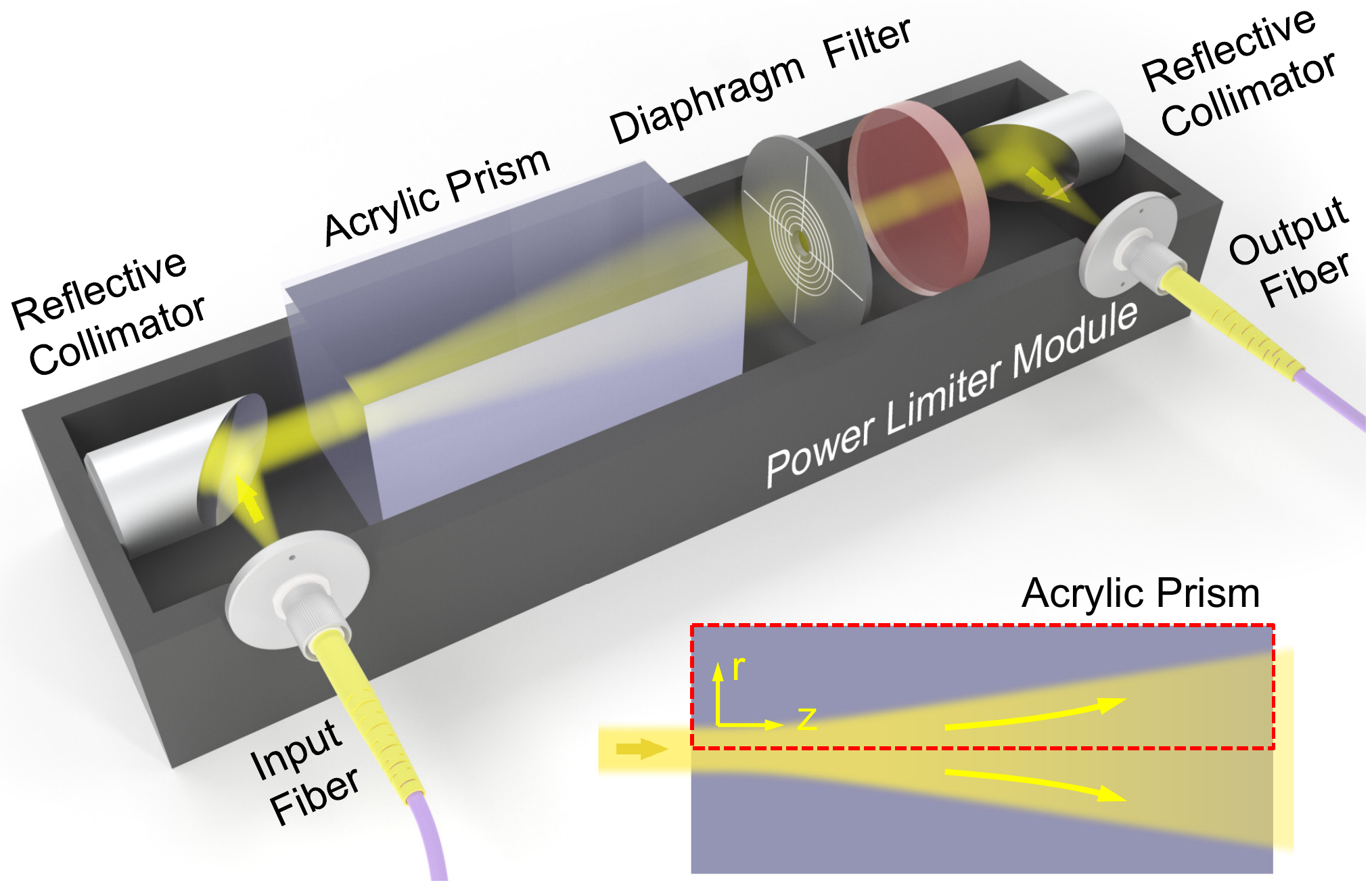}
\caption{Schematic of the power limiter design. An acrylic prism is used as the active medium. The input collimated Gaussian beam diverges due to the thermo-optical defocusing effect, when the absorbed energy introduces temperature gradients inside the prism. A diaphragm is placed after the prism to control the collectable optical power. The optical filter restricts the working wavelength range for security analysis. The inset is the top view of the acrylic prism and the diverged Gaussian beam. Owing to the isotropic nature of acrylic, both the optical and thermal responses are assumed to be axially symmetric along the optical axis.}
\label{fig1}
\end{figure}

Our power limiter design is shown in Fig.~\ref{fig1}. The input light and output light are collimated using a pair of fiber collimators. An acrylic prism is placed along the optical path as the core part of our proposal, whose negative thermo-optical coefficient (TOC) $ dn/dT $ is exploited, where $n$ is the refractive index and $T$ is the temperature. Noted here any material with negative TOC could be used with similar analysis. The absorption of input light generates a heat gradient inside the prism, which is then converted to a refractive index gradient accordingly. The negative TOC leads to a relatively smaller reflective index at the center of the prism, resulting in the whole optical architecture works as a concave lens and diverges the transmitting light, as shown in the inset of Fig.~\ref{fig1}. By adding a diaphragm with customizable width, the amount of output power can be suitably controlled. An optical filter is then introduced to restrict the working wavelength range of the device for security analysis, which will be discussed in detail in Section.~\ref{sect3}. We remark that all the components used here are cost-effective and commercially available.

\begin{figure}[tbp]
\centering
\includegraphics[width=1\columnwidth]{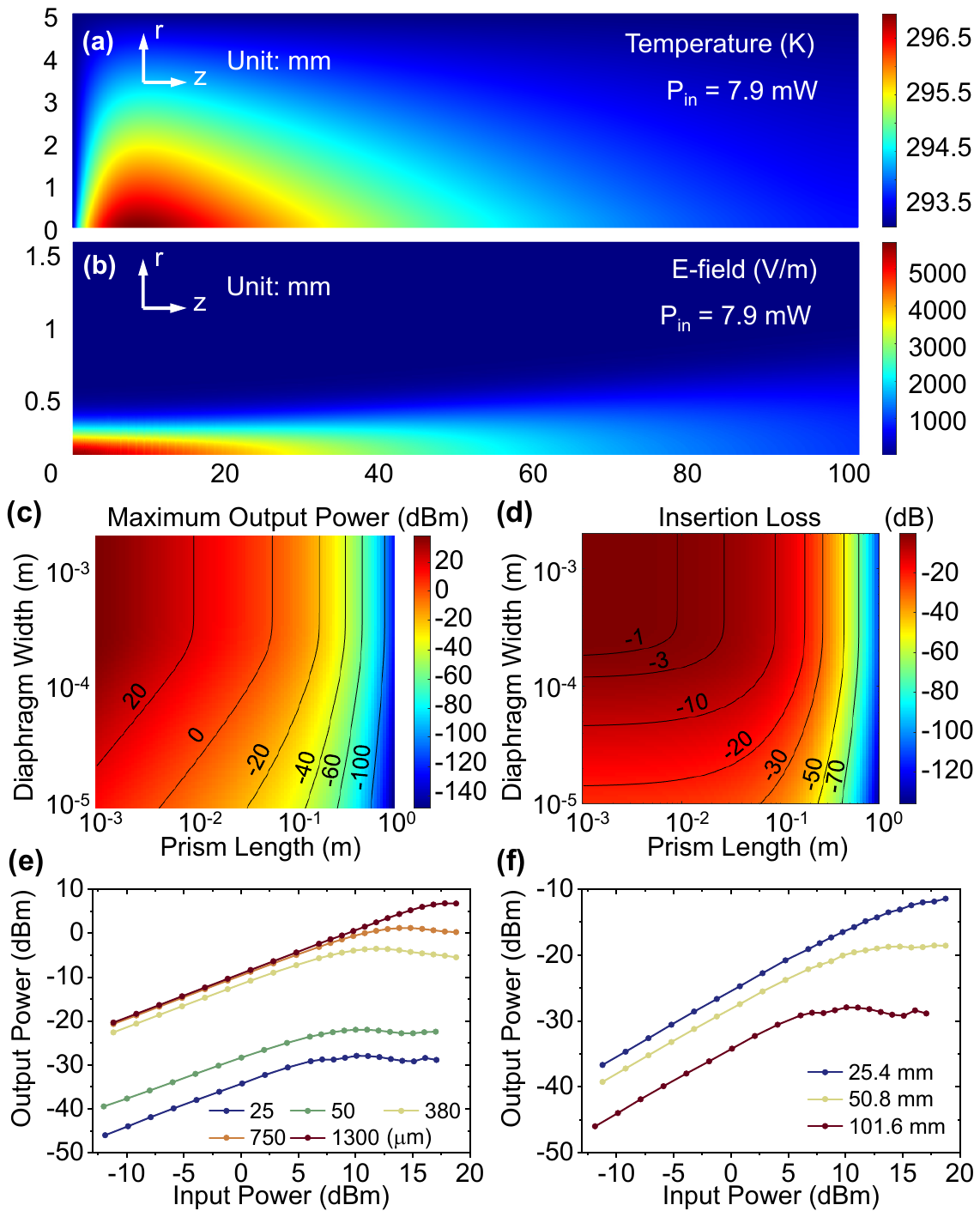}
\caption{Simulated (a) temperature and (b) E-field distribution in the region marked with the dashed box in Fig.~\ref{fig1} using a 2-dimensional model in COMSOL. The results clearly indicate the high-temperature gradient in the $r$-direction and the gradual divergence of Gaussian profile in the $r$-direction. Simulated (c) maximum output power and (d) insertion loss at different diaphragm width and prism length using Eq.~\ref{eq2}, where $\alpha = 25.95~{\rm m}^{-1}$ (measured~value),~TOC = $1.3\times10^{-4}~{\rm K}^{-1}$~\cite{zhang2006thermo}, $n = 1.47$~\cite{zhang2020complex}, $k = 0.19~ {\rm Wm^{-1}K^{-1}}$ \cite{rudtsch2004intercomparison},~$a = 0.14~{\rm mm}$.
(e) Experimental output-input power relationship at different diaphragm width with the same prism length of 101.6~mm. The results indicate that smaller the diaphragm width, lower the output power threshold. Also with a diaphragm width larger than the beam width, the insertion loss will remain minimum. (f) Experimental output-input power relationship at different prism length but with the same diaphragm width of 25~\textmu m. Longer prism length could provide lower output power threshold but the insertion loss will be higher. Both the simulation and experimental results confirmed the power limiting effect of our design and an adjustable power limiting threshold is feasible.}
\label{fig2}
\end{figure}

The mechanism of thermal optical defocusing and related power limiting phenomenon have been widely studied in both theory and experiments~\cite{smith_high-power_1977,leite_thermal_1967,derosa_fiber-optic_2003}. In our case, We first simulate the temperature and electric field distribution inside a 10~cm acrylic prism with 7.9~mW input power using COMSOL, whose results are shown in~\ref{fig2} (a) and (b). The simulation results indicate a distinct temperature distribution inside the medium, and a clear divergence of the light field. Towards a better quantitative understanding, we model our power limiter design by balancing the optical absorption and the heat transfer inside the prism under steady-state condition~\cite{smith1969thermal}:
\begin{equation}\label{eq1-1}
\alpha I = -\frac{k}{r}\frac{\partial}{\partial r} \left(r \frac{\partial T}{\partial r} \right),
\end{equation}
where $\alpha$ is the absorption coefficient of the material, $I$ represents the input light power density, $T$ is the temperature, and $k$ is the thermal conductivity. If we assume that the light propagates along the $z$-direction and follows a Gaussian profile, temperature gradient in the $z$-direction is negligible, and the radiative and convective heat transfer is minimal, the steady-state laser radiation intensity at position $(r,z)$ can be solved as~\cite{smith1969thermal}
\begin{equation}\label{eq2}
\begin{split}
I(r,z) =& I(r,0) \\
&\cdot \exp\Bigg[-\alpha z + \frac{\displaystyle\frac{\partial n}{\partial T} P_0 e^\frac{-r^2}{a^2} \left(z - \frac{1 - e^{-\alpha z}}{\alpha} \right)}{\pi kna^2}\Bigg],
\end{split}
\end{equation}
where the input intensity $ I(r,0)=\frac{P_0}{\pi a^2} e^{-r^2/a^2}$, $a$ is the radius where the light intensity drops to $1/e$ of its axial value, and $P_0$ is the incident laser power. 
The output optical power can be obtained by integrating the light intensity over a certain area which depends on the position (prism length) and the width of diaphragm.

The maximum output power (defined as power limiting threshold) and the insertion loss at different prism length and diaphragm width are shown in Fig.~\ref{fig2} (c) and (d). Since a larger prism length will lead to a greater photon absorption as well as a larger light divergence, a higher insertion loss and a smaller power limiting threshold can thus be expected. Likewise, a smaller diaphragm collects less photon energy, which also results in a higher insertion loss and a smaller power limiting threshold. Therefore, depending on the application, it is possible to choose a set of parameters that balance the insertion loss and power limiting threshold that meet system requirements. 

Note here that the Gaussian profile assumption only holds when the beam divergence is relatively small, thus the analytical model may only be able to provide a quick guidance for parameter selection. Hence, experiments are conducted to verify the feasibility of our proposal.

A proof-of-concept experiment is performed using a simplified version of Fig.~\ref{fig1}. A collimator is used for light coupling from single mode optical fiber to free-space. Here a transmissive collimator based on graded-index (GRIN) lens is used in the setup for feasibility demonstration, and it can be conveniently replaced by reflective collimators to ensure the proper functioning over a wide range of wavelengths for security reasons (See Section.\ref{sect3} for details). Then the Gaussian beam with a beam width of 0.4~mm is directed into the Acrylic Prism. Three acrylic prisms with lengths of 25.4, 50.8, and 101.6~mm are tested. The output light will then be collected after the diaphragm. Diaphragm width of 25, 50, 380, 750 and 1300~\textmu m are used in our experiment. Fig.~\ref{fig2} (e) shows the measured output-input relationship at different diaphragm width and the same prism length of 101.6~mm, while Fig.~\ref{fig2} (f) shows the result at different prism length with the same diaphragm width of 25~\textmu m. The results clearly show the power limiting effect in various conditions. The output power linearly increases with the input power at low power region. As the input power further increases, the output power will increase slowly, and finally be limited to a certain threshold. 

Besides, the experimental results verified that the power limiting feature of our proposal can be readily adjusted by modifying the prism length and diaphragm width. Among all of our system configurations, the lowest power limiting threshold of -27.9~dBm is measured, with a insertion loss of -34.0~dB, when a 101.6~mm prism and 25~\textmu m diaphragm are chosen. Similarly, a lower insertion loss of -5.1~dB can be obtained, together with a 10.3~dBm output power limiting threshold, when a 50.8~mm prism and 750~\textmu m diaphragm are used. For different applications, one can expect different requirements for power limiting device. For example, for protecting transmitters against THA, the insertion loss of the power limiter is less concerned since we can always adjust the optical attenuators to generate expected quantum states. While in order to protect receivers from bright illumination attacks, the insertion loss of the device can be a critical factor to system performance. Thus, we would imagine customised power limiter configurations for different application scenarios.

\section{Robustness against potential implementation loopholes}\label{sect3}

The above analyses so far only show the feasibility of the proposed power limiter under a steady-state condition. Below, we analyze the robustness of the proposed device against potential implementation loopholes that could happen via the variation of standard optical properties.

\begin{figure}[tbp]
\centering
\includegraphics[width=1\columnwidth]{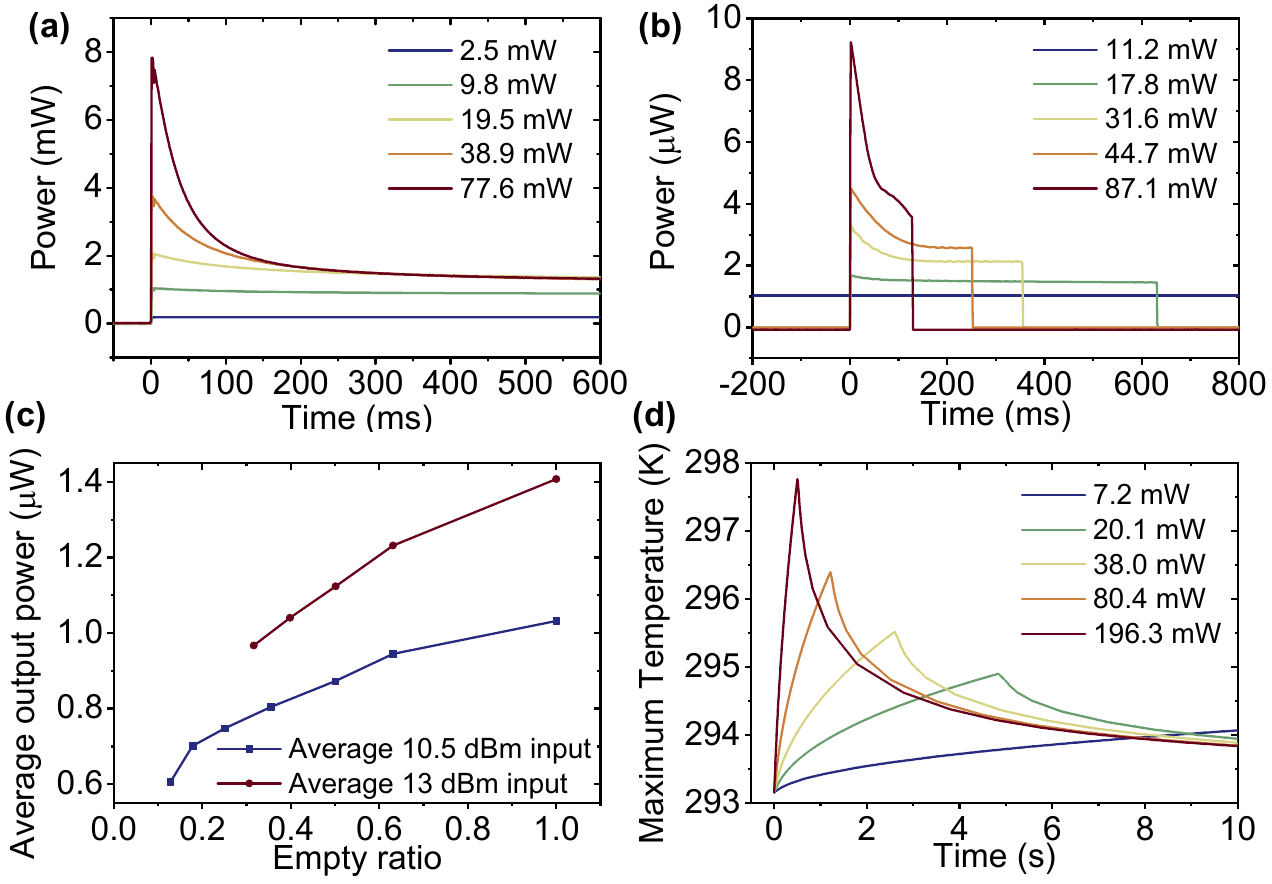}
\caption{(a) Experimental output response of the power limiter at different input optical power. An 101.6~mm prism is used with a 750~\textmu m diaphragm. The power limiting effect has a long settling time of around 300~ms. (b) Experimental output response of the power limiter with constant-energy pulse input. An 101.6~mm prism is used with a 25~\textmu m diaphragm. The peak power and duty cycle are selected to maintain the same energy per pulse or the same average input power of 10.5~dBm and 13~dBm. (c) The corresponding average output power at different duty cycle which shows that the maximum output power occurs at duty cycle of 1, i.e. CW input. (d) The COMSOL simulated maximum temperature inside the prism with constant average input power of 20~mW. The result shows that higher peak power heats the prism faster and reaches a higher temperature. Thus, a higher thermo-optical effect can be expected.}
\label{fig3}
\end{figure}

One important consideration is the finite response time of the proposed device. To investigate this property, we install an electronic variable optical attenuator (EVOA) after a continuous-wave (CW) laser source to create a laser pulse with relatively long pulse width and measure the output response of the power limiter. The experimental results are shown in Fig.~\ref{fig3} (a), where a 101.6~mm prism is used with a 750~\textmu m diaphragm. The settling time of our power limiter is measured to be 300~ms. We observed that the peak output power close to the starting time can be a few times higher than the steady-state output power (which happens after about 300~ms). Crucially, this suggests that one could exploit the finite response time of the power limiter to breach the desired energy threshold.

However, as we will show in Section. \ref{sect4}. A, the information leakage due to THAs can, in fact, be bounded using only the average energy constraint (integrated over the finite response time); it is not necessary to bound the maximum (peak) energy for security. Thus, in the experiment, we study the average output optical power at constant-energy pulse input but with different duty cycles. The time domain results are shown in Fig.~\ref{fig3} (b), where a 101.6~mm prism is used with a 25~\textmu m diaphragm. The input laser pulse is modulated at 1~Hz frequency with average input power of 10.5~dBm and 13~dBm. The corresponding average output power is shown in Fig.~\ref{fig3} (c). The results indicate that the average output power is higher at a larger duty cycle. The maximum appears at duty cycle equals 1, i.e. CW light input. In other words, given fixed average input power, CW input will give the largest averaged output power, where Eve is getting the most amount of information about the transmitter. As such, we will be using the power limiting threshold obtained under the CW Trojan horse input assumption for THA analysis; see Section. \ref{sect4}. A.
To explain this effect, we study the temperature response inside the medium under constant-energy pulse input with different peak power and different duty cycles using COMSOL. The results are shown in Fig.~\ref{fig3} (d). The simulation results indicate that a higher input peak power will lead to a higher maximum temperature, even with the same amount of average power. Therefore, a higher refractive index gradient and larger divergence of input laser are expected with a higher instantaneous power of the input light, leading to a larger thermo-optical defocusing effect and consequently a lower output power.

Another possible attack is to try to change the power limiting threshold by varying the wavelength of the incoming light. This could allow Eve to send in brighter light pulses with a different wavelength. To investigate the possibility of such an attack, we analyze how different input wavelength could affect the TOC and heat generation of the power limiter device. 

\begin{figure}[tbp]
\centering
\includegraphics[width=1\columnwidth]{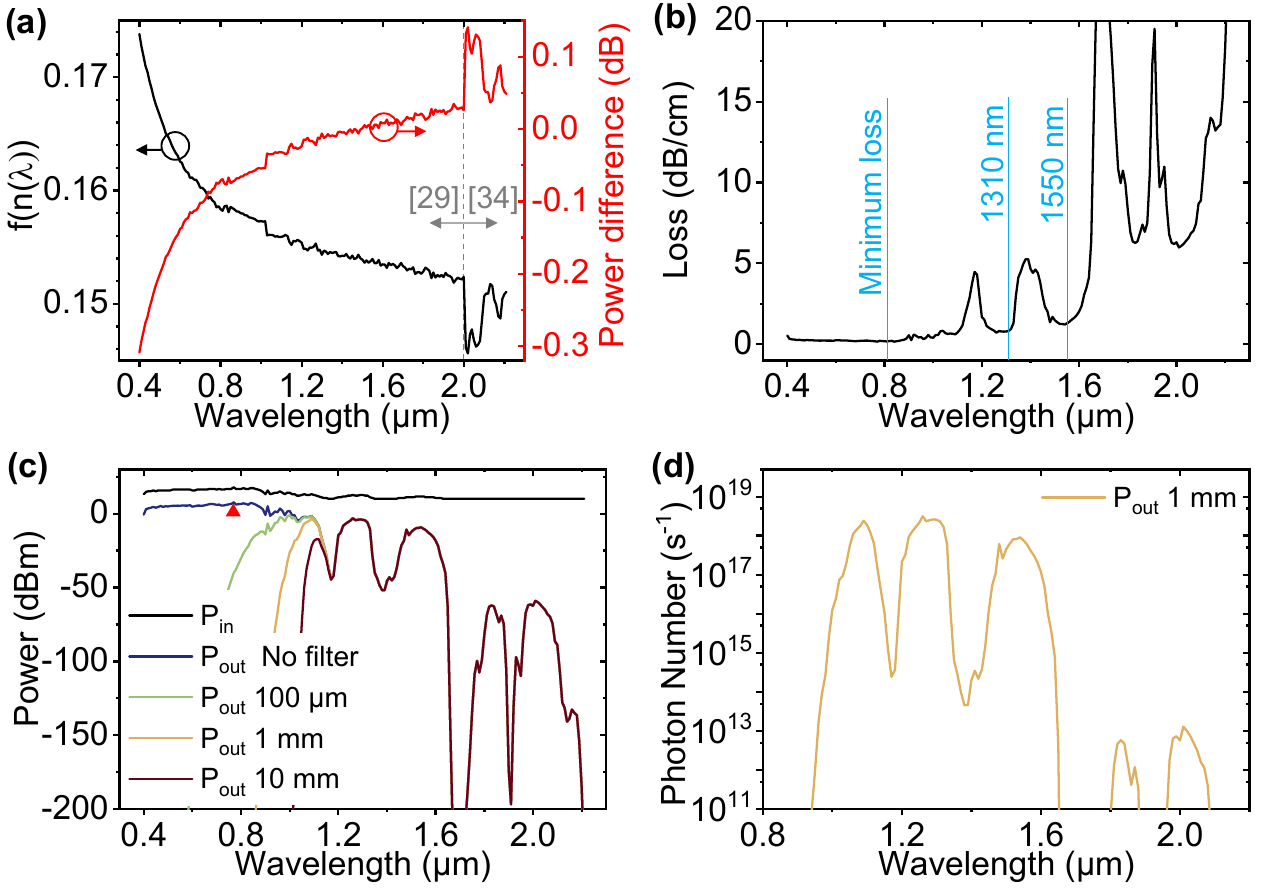}
\caption{(a) Calculated $f(n(\lambda))$ from Eq.~\ref{eq3} and the output power threshold difference caused by the corresponding TOC change, as referenced to the value at 1550~nm. (b) Absorption loss spectrum of acrylic, taken from Ref.~\cite{zhang2020complex,zhang2020complex2}. The loss at 1310~nm and 1550~nm communication bands are marked. The minimum loss is about 0.15~dB/cm at visible wavelength. (c) The maximum output power of the power limiter and the corresponding input power with silicon absorber (visible light absorbing filter) of different lengths. The silicon absorber significantly reduces the power threshold at the visible wavelength. (d) The total maximum output photon number per second calculated from (c) with the silicon absorber thickness of 1~mm.}
\label{fig4}
\end{figure}

Generally, the TOC can be modeled by~\cite{soave2009refractive,qiu_synthesis_2009} 
\begin{equation}\label{eq3}
{\rm TOC} = \frac{\mathrm{d} n}{\mathrm{d} T} = f(n(\lambda)) \left(\Phi-\beta\right),
\end{equation}
where $f(n(\lambda))$ is defined as $(n^2-1)(n^2+2)/(6n)$, $n$ is the reflective index, $\lambda$ is the wavelength of input light, $\Phi$ is the electronic polarizability and $\beta$ is the volumetric expansion coefficient. In most polymers, the volumetric expansion coefficient is more dominant, i.e. $\Phi\ll\beta$, and hence the overall TOC is typically negative~\cite{soave2009refractive}. More importantly, notice that the volumetric expansion coefficient is physically independent of the wavelength. As such, the wavelength dependency of TOC is only related to $f(n(\lambda))$. The $f(n(\lambda))$ for acrylic as the function of wavelength is shown in Fig.~\ref{fig4} (a). The corresponding TOC change will introduce a small difference in the output power threshold calculation, as referenced to the power at 1550~nm, which is shown as the red curve in Fig.~\ref{fig4} (a).

As for heat generation, it is related to the absorption loss of the material. A lower loss indicates less energy converted from the optical energy to heat energy, thereby resulting in a lower temperature gradient and a higher power limiting threshold. Based on this, a spectral filter with a large power handling capability can be applied to limit the transmission spectrum of the device; in which case the peak power (over the transmitted spectrum) is considered for the security analysis. 

Considering optical fiber-based applications at 1550 nm, the optical fiber itself is, in fact, a bandpass filter for about 300-2100 nm wavelength, beyond which the transmission loss is higher than 100~dB/km~\cite{agrawal2000nonlinear}. Thus, by applying a secure fiber with adequate length, light beyond this wavelength range can be suppressed to a negligible level. In this way, it is effective to only consider the wavelength dependency feature within this band.

For the material of our power limiter, acrylic, its absorption loss spectrum is shown in
Fig.~\ref{fig4} (b) with some low loss bands marked~\cite{zhang2020complex,zhang2020complex2}. The loss is about 1.29~dB/cm at 1550~nm and 0.82~dB/cm at 1310~nm, which are standard communication bands. The loss below 1100~nm is even lower. The minimum occurs at about 800~nm with 0.15~dB/cm loss. Based on the absorption spectrum and considering a prism length of 10~cm with 750~\textmu m diaphragm, the maximum output power spectrum and the corresponding input power is calculated based on Eq.~\ref{eq2}, as shown in Fig.~\ref{fig4} (c). The power threshold below 1100~nm is about 11~dB higher than the 1310~nm band and more than 17~dB higher than the 1550~nm band. Although a pessimistic power bound of about 8~dBm can be set and used as the system power bound (marked as a red triangle), it is better to use an optical filter to block the light below 1100~nm wavelength. The silicon absorber can be a good candidate, which provides a stable and robust filtering performance. By adding a thin layer of silicon sheet after the power limiter, the output power can be significantly suppressed. As shown in Fig.~\ref{fig4} (c), with only about 1~mm thick silicon, the maximum output power shifts back to the communication bands. The maximum output photon number per second can be further calculated, as shown in Fig.~\ref{fig4} (d). The maximum output photon number per second appears at 1260~nm wavelength with a photon energy of $1.58 \times 10^{-19}~J$; as such, this wavelength is considered in the security analysis using the worst-case approach.

Similarly, for other degrees of freedom, e.g. the state of polarization, the polymer acrylic used in our design inherently possesses isotropic behavior. Thus, by nature, it will not introduce any birefringence related changes and is independent of the thermo-optical effect, which avoids introducing related loopholes to the system. 

Another consideration is laser damage attacks~\cite{bugge2014laser,makarov_creation_2016,huang_laser-damage_2020}. 
Preliminary simulations indicate that the acrylic prism could be damaged with only about 400~mW of input power~\cite{m2014hole,berrie1980drilling}. 
If the acrylic is damaged or burnt, the thermal defocusing effect is not applicable anymore and the light might be collected by the output collimator directly. Consequently, the power limiting effect may not hold. However, this issue can be resolved by replacing the crossing-through prism with a total internal reflection structure, where the input beam is non-coaxial with the output. In this way, any damage to the material will not weaken the robustness of our proposal; instead the device works as an optical fuse to permanently block the optical path. 


\section{Quantum signal integrity}\label{sect3.5}

To determine if the proposed power limiter is useful for practical use, it is also important to study if the quantum signals will be disturbed when passing through the device. To this end, experiments based on time-bin (intensity), phase, and polarization encoding are implemented to see whether the quantum signal integrity will be affected.

Here, we study the QBER of the system, which is defined as the number of errors ($N_{error}$) over the total number of detection counts ($N_{correct} + N_{error}$),
\begin{equation}\label{eq_qber}
{\rm QBER} = \frac{N_{error}}{N_{correct} + N_{error}}.
\end{equation} In addition, given that the detector is well characterised (e.g., its background noise and single-photon efficiency are known), we may further write ${\rm QBER}= {\rm QBER}_{opt} + {\rm QBER}_{det}$,
where QBER$_{opt}$ comes from quantum optical imperfections (e.g., imperfect state preparation, optical misalignment, etc) and QBER$_{det}$ comes from the detector dark counts. Here, as mentioned above, our main focus is the QBER$_{opt}$ for intensity, phase and polarization encoding schemes, which represent three of the most popular choices for QKD in practice. 

\begin{figure}[tbp]
\centering
\includegraphics[width=1\columnwidth]{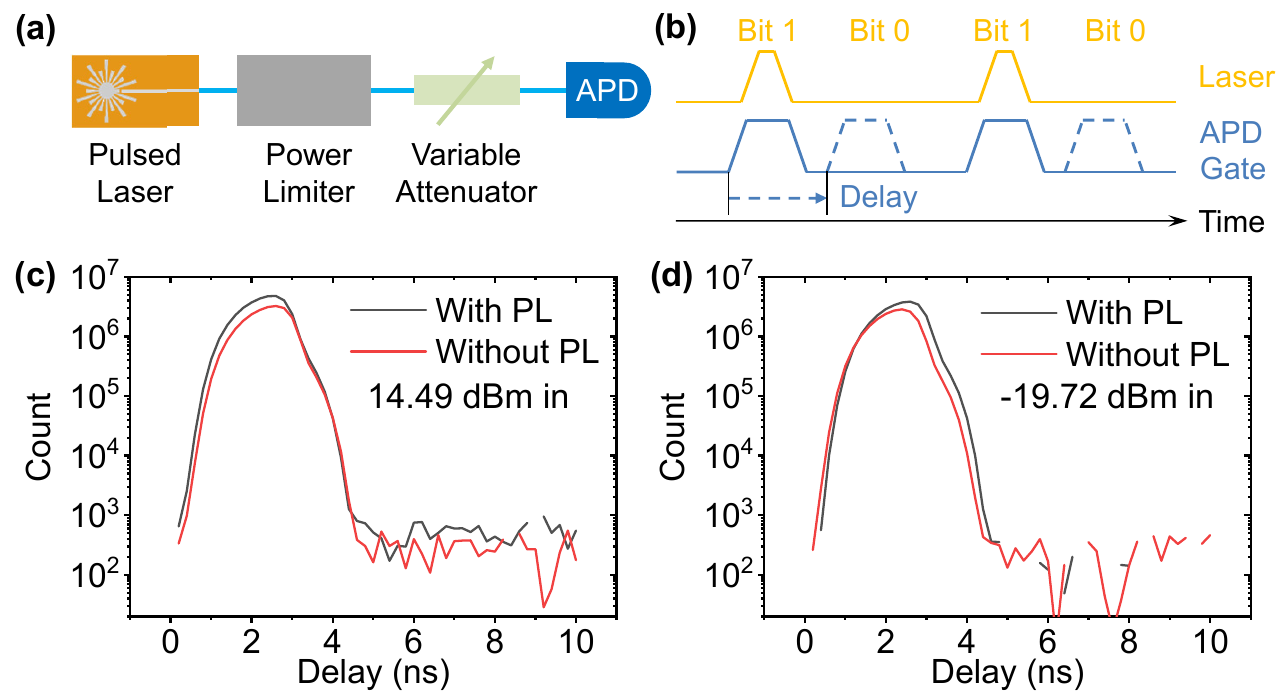}
\caption{(a) Experimental setup for QBER measurement of a time-bin encoding scheme (which uses on-off keying). An APD is used to measure the extinction ratio of a pulsed laser with and without the power limiter. (b) Schematic of the signal control for the pulsed laser and the APD gate signal. (c,d) The count value as a function of gate signal delay with and without power limiter (PL). (c) The case for 14.49 dBm input power, which is close to the power limiting threshold. (d) The case for -19.72 dBm input power. The discontinuity of the curve is because of the negative count values obtained due to the statistical noise. }
\label{fig_qber_i}
\end{figure}

The QBER of intensity or time-bin encoding scheme is measured first. As shown in Fig.~\ref{fig_qber_i}~(a), the intensity extinction ratio of a pulsed laser is measured to infer the QBER. The pulsed laser is attenuated to about 0.1 photon per pulse and measured by an avalanche photodiode (APD) operating in the Geiger mode (gated). The laser pulse has a repetition frequency of 100~MHz and a pulse width of 400~ps. The APD has a gate width of 1~ns. The delay on the APD gate signal is scanned to cover both the laser pulse (bit 1) and dark region (bit 0). The dark counts here are subtracted after the data acquisition for an accurate extinction ratio measurement of the optical pulse. The power limiter used here has a length of 101.6~mm and a diaphragm width of 750~\textmu m. An average input power of 14.49 dBm and -19.72 dBm are tested to demonstrate the cases when the input power is close to and far below the power limiting threshold, respectively. The schematic of the signal controls are shown in  Fig.~\ref{fig_qber_i}~(b). The resulting counts as a function of delay is shown in Fig.~\ref{fig_qber_i}~(c) and (d). For the input power and the cases with and without the power limiter, the resulting extinction ratios are all above 35 dB, indicating a QBER of less than 0.032\%. Therefore, we conclude that the introduction of the proposed power limiter will not introduce any significant noise to QKD systems based on time-bin encoding.

\begin{figure}[tbp]
\centering
\includegraphics[width=1\columnwidth]{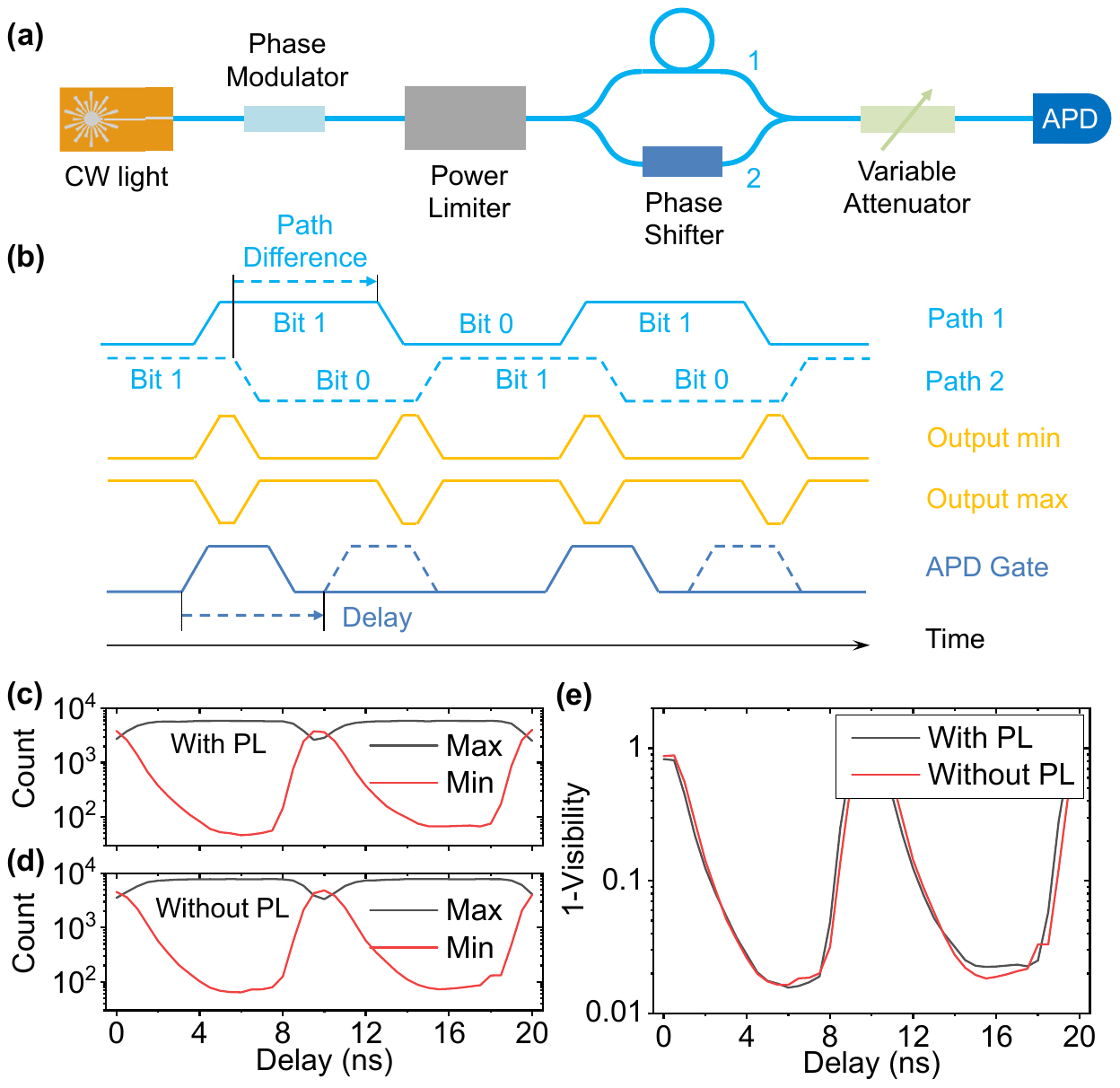}
\caption{(a) Experimental setup for the QBER measurement of a typical phase encoding scheme. The input CW light is modulated with a phase modulator in 0 and $\pi$ phase and decoded with an AMZI. The output light is attenuated to single-photon level and measured with an APD. (b) Schematics of the waveform for the modulated phase in both the path of AMZI and the outputs from AMZI with different phase shifter setting resulting in minimum power and maximum power. Similar to the time-bin (intensity) encoding scheme, the APD gate signal delay is scanned over the whole period. (c,d) The count value with minimum and maximum phase shifter setting as a function of gate signal delay with and without the power limiter. (e) The calculated interference visibility as a function of delay with and without the power limiter.}
\label{fig_qber_p}
\end{figure}

For the phase encoding scheme, the experimental setup is shown in Fig.~\ref{fig_qber_p}~(a). The input CW laser is modulated using a phase modulator switching between 0 and $\pi$ phase with 50~MHz frequency. The laser output power is 10.28 dBm. Schematics of the signal controls are shown in Fig.~\ref{fig_qber_p}~(b). The modulated signal is then decoded using an asymmetric Mach-Zehender interferometer (AMZI) with a path delay of around 10~ns. Moreover, a phase shifter is added in one of the paths of the AMZI to lock the relative phase. As such, the interference visibility as well as the QBER can be obtained. Finally the output is attenuated to 0.1 photon per gate and measured by an APD. The counts as a function of delay are shown in Fig.~\ref{fig_qber_p}~(c) and (d), which corresponds to the case with and without power limiter installed, respectively. The interference visibility $V$ is shown in Fig.~\ref{fig_qber_p}~(e) as $(1-V)$ for a clear view. The maximum visibility with and without power limiter are 0.9844 and 0.9836, corresponding to a QBER of 0.78\% and 0.82\%, respectively. Thus like in the case of time-bin encoding, we conclude that the proposed power limiter device is also suitable for phase-encoding QKD systems.

\begin{figure}[tbp]
\centering
\includegraphics[width=1\columnwidth]{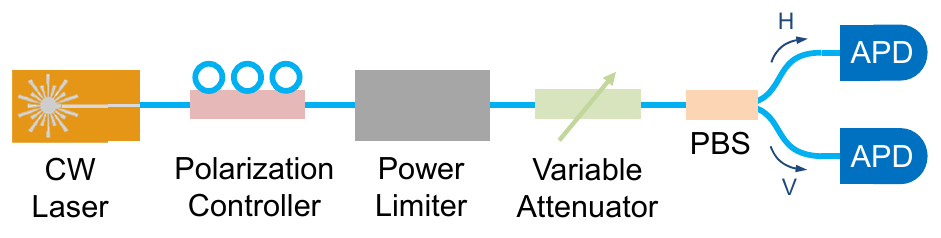}
\caption{(a) Experimental setup for the QBER measurement of the polarization encoding scheme. The polarization of a attenuated CW laser is manually tuned to match the polarization of a polarization beam splitter (PBS). The output from the PBS is measured by two APDs.}
\label{fig_qber_pol}
\end{figure}

Finally, we study the impact of the device on polarization encoding. The experimental setup is shown in Fig.~\ref{fig_qber_pol}, where a CW laser with an output power of 11.41 dBm is used and polarization is manually tuned with a polarization controller. The attenuated output goes through a polarization beam splitter (PBS) and the outputs are measured by two APDs. The polarization extinction ratio is calculated from the ratio between the two APD counts. The result shows a polarization extinction ratio of 30.1 dB and 32.6 dB for the case with and without the power limiter, corresponding to QBERs of 0.098\% and 0.055\%, respectively. This clearly shows that the power limiter will not significantly disturb the state of polarisation of the photon.

All in all, we experimentally confirmed that our power limiter device does not introduce any significant noise (in terms of the QBER) to standard QKD systems based on time, phase, and polarization encoding schemes. However, it should be noted that the power limiter does introduce extra losses (insertion loss) to the signal so the photon collection efficiency would decrease when it is deployed on the receiver side. In our experiment, a minimum insertion loss of -5.1~dB is measured, which is equivalent to an transmission efficiency of around 31\%, or a transmission distance of 25.5~km(assuming single mode fiber with a transmission loss of 0.2~dB/km). We note that this issue could be mitigated by using materials with higher TOC values so that smaller amount of light absorption is required to trigger the power limiting effect.

\section{Applications and countermeasures}\label{sect4}
\subsection{Security against THAs}\label{subsect4}
As an application of our proposed power limiter, we consider a phase-encoding MDI QKD protocol~\cite{Lo2012, Braunstein2012,tamaki2012phase} with energy constrained THAs. A schematic of our system is shown in Fig.~\ref{fig5}, where Alice and Bob are distant quantum transmitters and supposed to prepare the required phase-encoding coherent states, then send them to Charlie for Bell-state measurement. The protocol is outlined below: 

\begin{figure}[tbp]
\centering
\includegraphics[width=1\columnwidth]{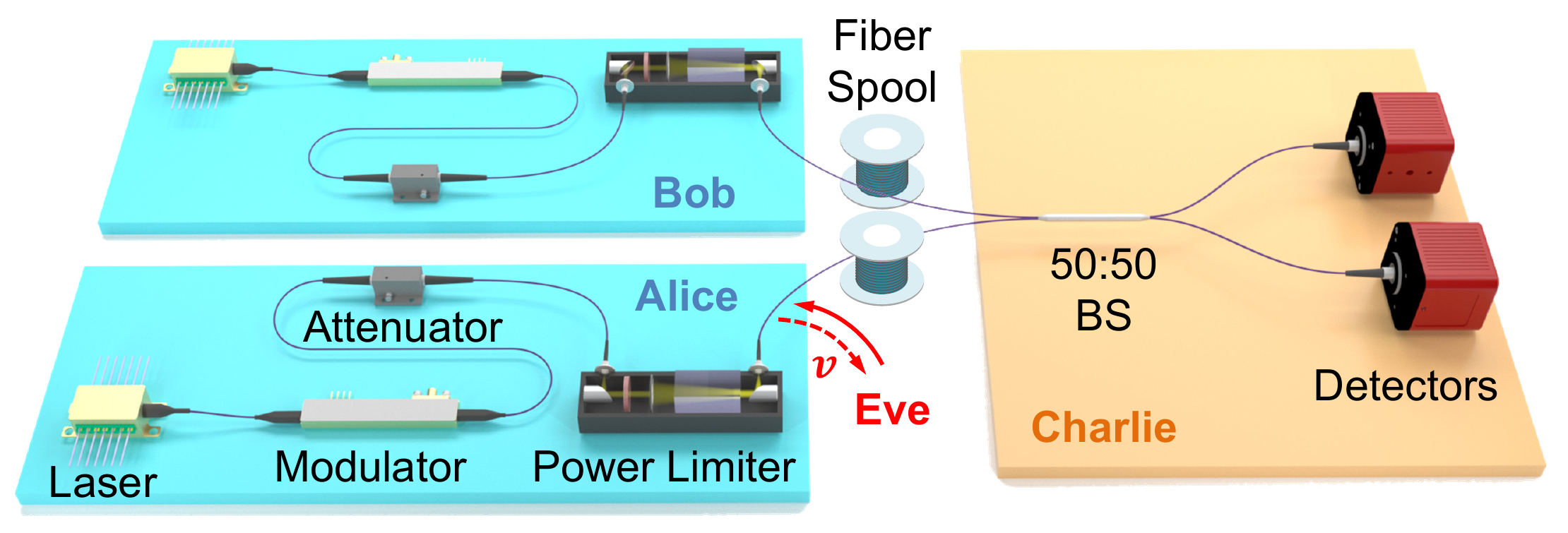}
\caption{Schematic of the phase-encoding MDI QKD system with the power limiter installed. Alice and Bob are the users preparing the phase-encoding coherent states using their lasers, modulators, and attenuators. The prepared states are sent to Charlie for Bell state measurement. The distance between Alice and Bob is contributed by the two fiber spools combined. The Trojan horse attack from Eve could provide her with a maximum $\nu$ Trojan horse photon, which is taken into consideration for secure key rate calculation.}
\label{fig5}
\end{figure}

Alice and Bob randomly prepare one of the four coherent states $\{\ket{e^{ix\frac{\pi}{2}}\alpha}\}$ and $\{\ket{e^{iy\frac{\pi}{2}}\beta}\}$, where $x,y\in\{0,1,2,3\}$ are the classical information of Alice and Bob, respectively. Then Alice and Bob send the quantum states to Charlie via the quantum channel for Bell-state measurement. The distance between Alice and Bob is contributed by the two fiber spools combined. 

Charlie interferes the incoming states from Alice and Bob using a 50-50 beam-splitter and measures the outputs using single-photon detectors. Thereafter, he announces the measurement result $z\in\{L,R,\varnothing\}$ through the authenticated classical channel, which corresponds to the left detector clicks, the right detector clicks, and none of the detector clicks or both detectors click. Alice and Bob repeat the state preparation and measurement for $N$ rounds.

Upon receiving the Bell-state measurement results from Charlie, Alice and Bob only keep data of those rounds give $z = L, R$. Besides, Bob flips the value of $y$ if $z=R$. Alice and Bob then obtain the statistics of all the state combinations of Alice and Bob, conditioned on $z=L,R$. Particularly, for rounds with $x,y=0,2$, Alice and Bob keep the data for extracting the secret keys. 

Alice and Bob then implement parameter estimation and apply error correction and privacy amplification thereafter to extract a pair of identical and secure keys.

To take THAs into consideration, different models for Trojan horse states have been proposed. For example, in Ref. \cite{lucamarini_practical_2015}, the Trojan horse state is modeled as a pure coherent state with a fixed phase and intensity. However, this model might be too restrictive as Eve can send other states. In practice, she could send a mixture of coherent states with different intensities or other states that could potentially leak more information. Another model that can address potential THAs is presented in Refs. \cite{pereira2019leaky, pereira2020correlated}. There, the non-vacuum component of the Trojan horse state is modeled by an arbitrary state that lives outside the qubit space in which the legitimate parties encode the information. While this model is very general and could take into accounts of any source side-channels, the resulting bound can be overly pessimistic, since in the worst-case scenario the leakages might correspond to orthogonal quantum states and hence would leak full information about the modulation (key information).

In our analysis, we take the intermediate step and allow Eve to send any Trojan horse state in a given optical mode.  However, because the modulators in Alice's and Bob's labs are trusted, the resultant Trojan horse states will not be orthogonal after the modulation. As such, the THA will not leak complete information about Alice's and Bob's key information. Without loss of generality, the Trojan horse state can be written as
\begin{equation}\label{eq4}
\ket{\xi}=\sum_{n,m}{c_{nm}\ket{n}\ket{m}}\ket{\mathcal{E}_{nm}},
\end{equation}
where $\ket{n},\ket{m}$ are the Fock states injected into Alice's and Bob's apparatus, respectively. $\ket{\mathcal{E}_{nm}}$ is an ancilla that is kept in Eve's lab. The coefficients $c_{nm}$ are the quantum amplitudes of the Fock states. Note that the state of the form \eqref{eq4} includes Trojan horses that are mixed (after tracing out Eve's ancilla) and may even be entangled. 

The states $\ket{n}$ and $\ket{m}$ will accumulate some phases introduced by Alice's and Bob's modulators and hence they would leak some information about $x$ and $y$. On the other hand, the states $\ket{\mathcal{E}_{nm}}$ will not accumulate any phase since it is kept in Eve's lab. After gathering the modulation information from the modulators, the output THA state thus with the form
\begin{equation}\label{eq5}
\ket{\xi'_{xy}}=\sum_{n,m}{c_{nm}e^{i(nx+my)\frac{\pi}{2}}\ket{n}\ket{m}}\ket{\mathcal{E}_{nm}},
\end{equation}

Both the quantum states prepared by Alice and Bob and the THA state will be sent to Charlie via the quantum channel. Thus, the untrusted measurement can be modeled by a quantum-to-classical map, which can be described by an isometry $\mathcal{U}$ (with an appropriate purification):
\begin{align}\label{eq6}
\ket{\phi_{xy}}&=\ket{e^{ix\frac{\pi}{2}}\alpha}\ket{e^{iy\frac{\pi}{2}}\beta}\ket{\xi'_{xy}}\nonumber\\
&\stackrel{\mathcal{U}}{\longrightarrow}\sum_{z}{\ket{\mathrm{e}^z_{xy}}\ket{z}}.
\end{align}
Therefore, given the fact that the Gram matrix $G$ based on Eq. (\ref{eq6}) is positive semi-definite and linearly constrained, a tight upper bound of the phase error rate can be obtained by solving the dual problem of a semi-definite program (SDP), similar to the security analysis presented in Ref.~\cite{wang_characterising_2019,primaatmaja2019versatile}.

The asymptotic secret key rate can thus be obtained using the so-called Shor-Preskill key rate formula~\cite{Shor2000}:
\begin{equation}\label{eq8}
R \geq \max\{0,P_{\rm pass} [1-h_2(e_{\rm ph})-h_2(e_{\rm bit})] \},
\end{equation}
where $e_{\rm bit}(e_{\rm ph})$ is the bit (phase) error rate of the system, $P_{\rm pass}$ represents the probability of successful Bell state measurement when Alice and Bob choose the key generation basis, and $h_2(\cdot)$ is the binary entropy function. A detailed security analysis is given in the supplemental material. To restrict information leakage, the mean photon-number, $\nu$, of the THA state should be much less than one. This requirement can be achieved using the proposed power limiter together with an optical attenuator. These devices can be readily implemented in standard quantum transmitters as shown in Fig.~\ref{fig5}.

To be more specific, based on the power limiting threshold obtained in Section.~\ref{sect3}, 
a maximum photon number of injected eavesdropping light can be strictly constrained by the proposed optical power limiter. Then, the injected light will go through the attenuator twice before being collected by Eve, while the quantum state for QKD has just been attenuated once. Consider a QKD system working at a frequency of 1~GHz, with a power limiting threshold of 1~mW  and an ideal phase modulator that does not introduce any extra insertion loss. In this case, an attenuation of 69~dB is sufficient to guarantee an average energy output of $\nu=10^{-7}$. At the same time, the laser output can be adjusted to optimize the intensity $\mu$ for QKD, where an averaged optical power of 23~\textmu W can be used to generate quantum states with $\mu=0.0183$. This is similar to the optimized intensity for MDI QKD with a detector efficiency of $\eta_\text{det} = 85\%$, dark count rate $p_\text{dc} = 10^{-7}$, and a 100~km transmission distance. Comparing to Ref.~\cite{lucamarini_practical_2015} where the 12.8~W optical fiber damage threshold was used as the upper bound, the proposed power limiter could limit the power by 4 to 5 orders of magnitude lower. As a result, the requirement for attenuator and optical isolator is significantly reduced. Also, removing the need to have isolators could benefit future chip-based integration of such MDI QKD systems.

As mentioned, due to the finite response time of the power limiter, only the average power of the THA (instead of the maximum power of the THA pulses) can be bounded. To this end, we develop a general proof technique (see the supplemental material) that uses only the average photon number information of the THA. In particular, the security proof takes into account attacks where Eve employs a mixture of bright Trojan horse pulses with the vacuum (where the probability of sending a bright light is small enough such that the energy constraint is satisfied). As such, the proposed optical power limiter can be used to ensure that the assumptions of the security proof are enforced during the protocol.

\begin{figure}[tbp]
\centering
\includegraphics[width=1\columnwidth]{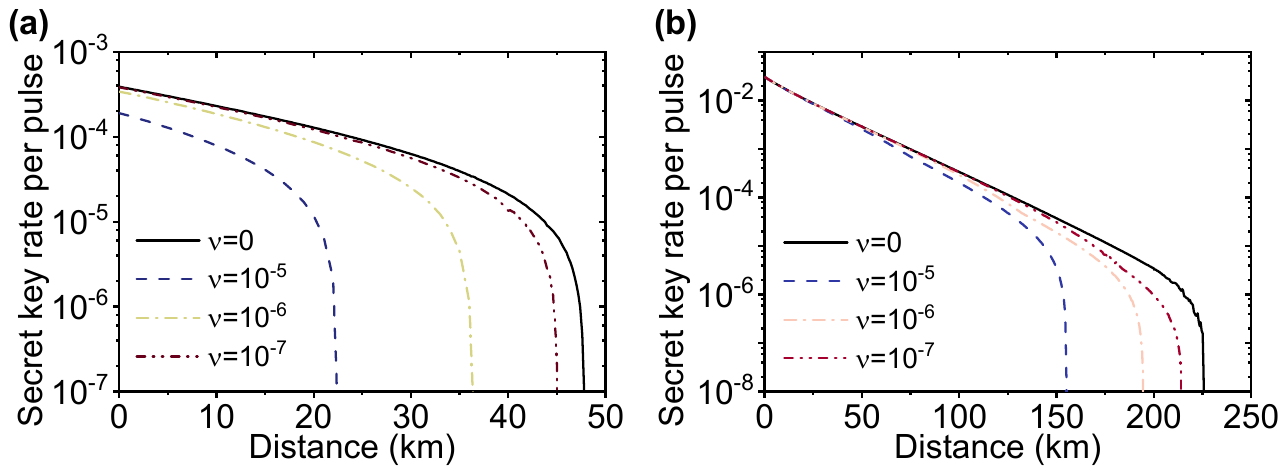}
\caption{Simulation for asymptotic key rate for phase-encoding MDI QKD under two set of parameters: (a) detector's efficiency $\eta_\text{det} = 10\%$, dark count rate $p_\text{dc} = 10^{-5}$, (b) detector's efficiency $\eta_\text{det} = 85\%$, dark count rate $p_\text{dc} = 10^{-7}$. Trojan horse photon number $\nu$ of $10^{-5}$, $10^{-6}$, $10^{-7}$ and 0 are shown. The output intensity $\mu$ of each transmitter is optimized for each distance to maximize the key rate.}
\label{fig6}
\end{figure}

To benchmark the performance of the protocol, we simulate the achievable asymptotic key rate with two sets of parameters: (1) detector's efficiency $\eta_\text{det} = 10\%$, dark count rate $p_\text{dc} = 10^{-5}$, (2) detector's efficiency $\eta_\text{det} = 85\%$, dark count rate $p_\text{dc} = 10^{-7}$. For both sets of parameters, misalignment error $e_{\text{ali}}$ is set to be $2\%$, and the transmission loss of optical fiber is set to be 0.2 dB/km. We also assume that the central node is equidistant to Alice and Bob and $\mu_A = \mu_B = \mu$, which has been optimized over the simulation. As for the THA intensity, we set $\nu_A=\nu_B=\nu$. The results of the simulation are shown in Fig.~\ref{fig6}. The results indicate Alice and Bob can get a promising key rate without being affected much by the Trojan horse attack if the energy of the THA is properly upper bounded.

\begin{figure}[tbp]
\centering
\includegraphics[width=1\columnwidth]{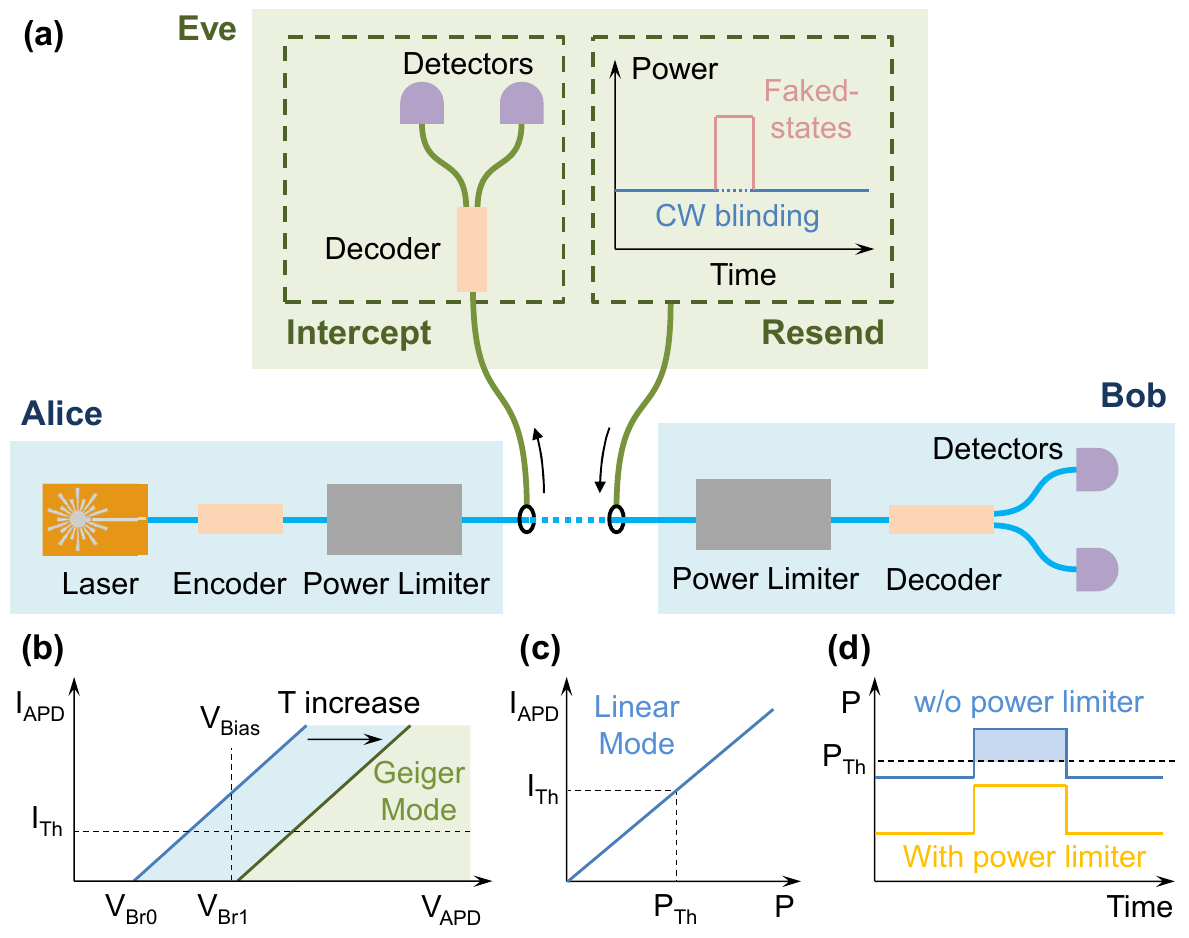}
\caption{(a) A schematic of the detector blinding attack proposed in Ref.~\cite{lydersen_thermal_2010}. (b) The current-voltage relationship of a typical APD. In normal working conditions, the breakdown voltage is $V_{Br0}$. A fixed bias voltage $V_{Bias}>V_{Br0}$ enables Geiger mode operation with single-photon sensitivity. The resulting output current above a threshold $I_{Th}$ is registered as a successful count. However, the breakdown voltage increases with the temperature $T$ of the device. Thus, it is possible to blind the detector by driving it into the linear mode ($V_{Bias}<V_{Br1}$), thus nullifying the single-photon sensitivity. (c) The current-input power relationship of an APD operating in linear mode. By controlling the input power below or above the power threshold $P_{Th}$, the detector could be controlled to register fake-states. (d) The input power on Bob's detector with and without a power limiter.}
\label{fig_blind}
\end{figure}

\subsection{Potential countermeasures}
\subsubsection{Bright illumination attacks}

Laser damage attacks are a particularly powerful class of bright illumination attacks. In Ref.~\cite{bugge2014laser,makarov_creation_2016,huang_laser-damage_2020}, it is shown that the detectors and optical components are prone to permanent changes and damages when Eve sends in a bright damaging laser with power in the order of Watts. This is crucial because the security of most QKD systems depend on the integrity of their devices---that they behave according to design specifications. 

Another class is detector blinding attacks~\cite{makarov2009controlling,lydersen2010hacking,yuan_resilience_2011}. By exploiting the implementation knowledge of single-photon detectors and the imperfect detector performances, Eve can send in a relatively strong eavesdropping light to change the working condition of the detector and get partial (or even full) control over the outcomes~\cite{makarov2009controlling,lydersen2010hacking,yuan_resilience_2011,lydersen_controlling_2011,qin_homodyne-detector-blinding_2018}. 

For illumination-related attacks, a common feature is that Eve must send in relatively bright light pulses. Hence, by restricting the input optical power using the proposed power limiter, it is expected that some of these attacks could be thwarted. To illustrate this possibility, we sketch out a method that could prevent the bright illumination attack presented in Ref.~\cite{lydersen_thermal_2010}; see Fig.~\ref{fig_blind} (a). To start with, we note that standard single-photon detectors based on APD typically require low-temperature operation to minimize the detectors' background noise, i.e., to limit the dark count rate. To cool the detectors, thermoelectric coolers (TECs) are used but these have limited cooling capacity. In Ref.~\cite{lydersen_thermal_2010}, it is shown that injection of bright light pulses can create a situation in which the generated heat fails to dissipate completely. This leads to the breakdown voltage of the APD going above the predetermined value, which consequently puts the detector into the \emph{linear} mode (instead of \emph{Geiger} mode); see Fig.~\ref{fig_blind} (b). In this case, the detector is no longer sensitive to single-photon input (i.e., \emph{blinded}) and Eve can manipulate its outcome by sending in a control light pulse superimposed on the bright light pulse, as depicted in Fig.~\ref{fig_blind} (c). 

According to Ref.~\cite{lydersen_thermal_2010}, a bright CW light with an optical power of around 10~mW is required to blind the commercial QKD detectors, and a control light pulse with a peak power of around 1 mW is sufficient to fully control the detector's outcome. If the power limiter is in place, as shown in Fig. 7 (a), the input light power can be limited below than this blinding threshold, which would prevent the temperature of detector from raising and hence the detector from being blinded (see Fig.~\ref{fig_blind} (d)). For example, we can use an acrylic prism with length of 50.8~mm and a diaphragm width of 380~\textmu m to provide a power limiting threshold of 6.03~dBm (with -6.02~dB insertion loss) to prevent such attack. It's important to note that in normal working conditions, quantum signals (i.e., optical signal with small energy levels) in principle will experience a small amount of loss while passing through the power limiter device, since the power limiting effect hasn't been triggered. As such, the introduction of our power limiter are not expected to greatly reduce the overall performance of practical QKD systems. As with our current design, we can expect a smaller insertion loss as well as a stronger power limiting effect, for example, by using material with a higher TOC.

\subsubsection{Plug-and-play QKD with untrusted light sources}\label{sect_twowaycomm}
Plug-and-play QKD is a two-way communication configuration~\cite{muller_plug_1997} that aims to simplify implementation requirements such as polarization compensation and reference frame calibration. This approach is especially useful for practical MDI QKD systems since it naturally guarantees near-perfect mode matching for the required two-photon interference~\cite{xu_measurement-device-independent_2015,tang2016experimental,wang_phase-reference-free_2015}.

\begin{figure}[tbp]
\centering
\includegraphics[width=0.7\columnwidth]{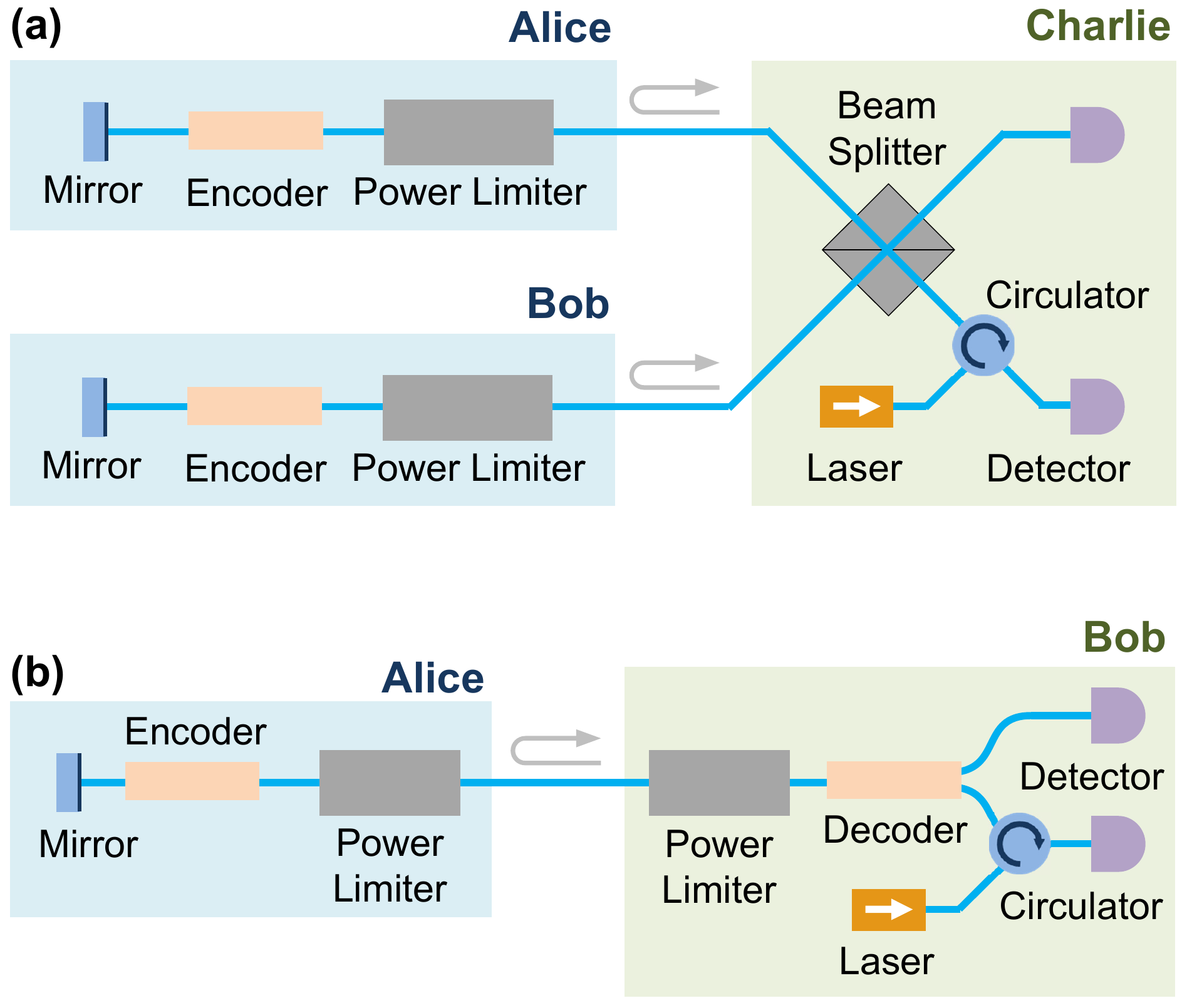}
\caption{Schematics of the power limiter used in (a) a plug-and-play MDI QKD system and (b) a plug-and-play BB84 QKD system.}
\label{fig_pnp}
\end{figure}

However, in using external (untrusted) light sources instead of trusted light sources, plug-and-play systems are prone to transmitter-based attacks~\cite{gisin_trojan-horse_2006,xu_experimental_2010,sun_passive_2011}. Again, the central issue here is that Eve can inject bright light pulses to break the working assumptions of QKD. To overcome this issue, one popular approach is to monitor the energy of the incoming light with a classical detector~\cite{zhao_quantum_2008,zhao_security_2010,xu_measurement-device-independent_2015}. However, it has been shown that such active monitoring methods are not entirely robust and the classical detectors can still be hacked by exploiting their electrical circuitry, e.g., see Refs.~\cite{makarov_creation_2016,sajeed_attacks_2015}.

In light of the above, it is thus interesting to explore alternative countermeasures that are based on passive devices instead of active devices such as detectors. To this end, we propose to replace (or augment) the active power monitoring device with a passive power limiter as shown in Fig.~\ref{fig_pnp}. Similar to the arguments provided in Section.~\ref{subsect4}, the power limiter would limit the energy of the outgoing light and hence Eve's knowledge about the key information as well; we leave a more careful security analysis to future work. 

In addition, it is worthwhile to add that existing methods to limit incoming light energy are typically based on isolators/circulators and laser damage threshold of devices~\cite{lucamarini_practical_2015,makarov_creation_2016,huang_laser-damage_2020}. These are however one-directional and add additional attenuation on the propagating direction of the eavesdropping light. In the case where quantum signal has the same propagation as Eve's light, i.e., plug-and-play QKD or quantum receivers to resist against bright-illumination attacks, they may either pessimistically estimate Eve's information---since the actual input power significantly deviates from the device damage threshold, which will be used for security analysis, or introduce large insertion loss so the system performance will be greatly affected.

As a comparison, the proposed power limiter is shown to be able to provide an adjustable power limiting threshold on the output optical power, and capable of protecting the system where the eavesdropping light and quantum signal have the same propagation direction.

\section{CONCLUSION}\label{sect5}

In this report, we have proposed and demonstrated a passive power limiter design based on the thermo-optical defocusing effect of an acrylic prism. By numerical simulations and the experimental demonstration, we rigorously studied the feasibility and performance of our power limiter design. In our experiment, the lowest optical power limiting threshold of -27.9~dBm with an insertion loss of -34.0~dB is measured. With a different setting, a low insertion loss of -5.1~dB is achieved with a 10.3~dBm power limiting threshold. The values are adjustable according to different system requirements. It is possible to further reduce the insertion loss at a certain power threshold by switching to a material with higher TOC values or/and reduce the beam width. Besides, our design possess desirable features like compactness, robustness, plus polarization and spectrum-dimension independence.

To illustrate the applicability of our proposed power limiter, we have quantitatively developed a general security analysis that allows for arbitrary of Trojan-horse states. By properly limiting the THA energy leakage in a MDI QKD system, a desirable secure key rate and transmission distance can be achieved. Moreover, based on the previous evidences, we remarked that the power limiter can be useful for deterring bright illumination attacks in a quantum cryptography system. We took the thermal CW-blinding attack on the APD detectors as an example, and show how the power limiter can be designed to prevent such an attack. We further discussed the possibility of using a power limiter to secure the plug-and-play QKD systems without active elements.

As demonstrated in our paper, by simply limiting the incoming/outgoing optical energy, a broad class of QKD protocols can be practically protected without introducing cumbersome device modification. Beyond these, one can also expect such a power-limiting device to find applications in securing semi-device-independent quantum protocols based on energy constraints~\cite{van2017semi,van_himbeeck_correlations_2019,avesani2020semi,rusca2019self_testing}, line-topology or ring-topology multiparty quantum communication systems~\cite{grice_two-party_2015,schmid_experimental_2005}. As such, we believe it will attract much interest and possess the potential to become a standard tool for quantum cryptography applications.

\begin{acknowledgments}
This research is supported by the National Research Foundation (NRF) Singapore, under its NRF Fellowship programme (NRFF11-2019-0001) and Quantum Engineering Programme 1.0 projects (QEP-P2, QEP-P3, and QEP-P8)
\end{acknowledgments}

\appendix

\section{Detailed security analysis}
In this supplemental document, we present the detailed security analysis of the phase-encoding MDI QKD protocol presented in the main text. To analyze the security of the protocol, we modify the security proof technique presented in Ref. \cite{primaatmaja2019versatile} to account for arbitrary Trojan horse attack (THA). Following the argument of Ref. \cite{primaatmaja2019versatile}, the untrusted measurement can be described by an isometry $\mathcal{U}$ acting on the effective signal state $\ket{\phi_{xy}}$ (which includes the Trojan horse photons)
\begin{equation}
    \ket{\phi_{xy}} \stackrel{\mathcal{U}}{\longrightarrow} \sum_z \ket{\rm{e}_{xy}^z} \ket{z}
\end{equation}
where $\ket{z}$ is the classical register that stores the outcome of the Bell state measurement and $\ket{\rm{e}_{xy}^z}$ is a sub-normalized state which stores Eve's quantum side information.

The Gram matrix associated to Eve's quantum side information, denoted by $G$, has to be positive semi-definite (PSD). Moreover, the observed statistics $\{P(z|x,y)\}_{x,y,z}$ put some constraints on $G$
\begin{equation}
    P(z|x,y)= \braket{\mathrm{e}^z_{xy}|\mathrm{e}^z_{xy}}
\end{equation}
Furthermore, since the measurement is described by an isometry, the inner product has to be conserved, i.e.
\begin{equation}
    \braket{\phi_{x'y'}|\phi_{xy}} = \sum_z \braket{\mathrm{e}^z_{x'y'}|\mathrm{e}^z_{xy}}
\end{equation}
Finally, the phase error rate, which quantifies how much information being leaked to Eve, is a linear function of the Gram matrix. Following the derivation in Appendix A of Ref. \cite{primaatmaja2019versatile}, the phase error rate is given by
\begin{equation}
\begin{split}
    \eph = \frac{1}{2} + \frac{1}{4 P_{\text{pass}}} \text{Re}\Bigg\{ &\braket{\mathrm{e}^{L}_{00}|\mathrm{e}^{L}_{22}} - \braket{\mathrm{e}^{R}_{00}|\mathrm{e}^{R}_{22}} \\ &- \braket{\mathrm{e}^{L}_{02}|\mathrm{e}^{L}_{20}} + \braket{\mathrm{e}^{R}_{02}|\mathrm{e}^{R}_{20}} \Bigg\}
\end{split}
\end{equation}
where
\begin{equation}
\begin{split}
    P_{\text{pass}} = \frac{1}{4}\Big[ &P(L|00) + P(R|00) + P(L|02) + P(R|02)  \\ & +P(L|20) + P(R|20) + P(L|22) + P(R|22) \Big]
\end{split}
\end{equation}
is the probability of successful Bell state measurement given Alice and Bob choose the key generation basis (i.e. when $x,y \in \{0,2\})$ which is observed directly in the experiment. The linearity of the constraints and the objective function allows us to use semi-definite programming (SDP) to find a tight bound on the phase error rate $\eph$.

For fixed observed statistics, the phase error rate $\eph$ depends only on the Gram matrix of the effective signal states $\{\ket{\phi_{xy}}\}_{x,y}$. Our goal is therefore to characterize the set of Gram matrices of the effective signal states subject to the constraint on the mean energy of the Trojan horse lights. Indeed, the main difference between our security analysis and the one presented in Ref. \cite{primaatmaja2019versatile} lies in the fact that $\braket{\phi_{x'y'}|\phi_{xy}}$ is not perfectly characterized, but it can be bounded due to the energy constraints on the Trojan horse state.

Assuming that Alice and Bob each modulate lights from a single mode, the most general form of Trojan horse state that Eve can send is given by
\begin{equation}
    \ket{\xi}_{a b e}=\sum_{n,m}{c_{nm}\ket{n}_{a}\ket{m}}_{b}\ket{\mathcal{E}_{nm}}_{e},
\end{equation}
where the registers $a,b,e$ are held in Alice's, Bob's, and Eve's lab respectively. The states $\ket{n}$ and $\ket{m}$ denote photon number states. In general, the Trojan horse can also be entangled to some ancillary system $e$ that is kept in Eve's lab which we denote by $\ket{\mathcal{E}_{nm}}$. In practice, only a fraction of the Trojan horse lights is reflected out to the quantum channel. The rest of the photons are lost (and, therefore, are inaccessible to Eve). In our model, we conservatively ignore the loss in the modulators. Finally, for the remainder of this section, we will omit the subscripts denoting the registers when there is no danger of ambiguity.

Now, Alice's and Bob's phase modulations can be described by the unitary operators
\begin{equation}
\begin{split}
    \hat{U}_x = e^{i\frac{\pi}{2} x \hat{n}}\\
    \hat{U}_y = e^{i\frac{\pi}{2} y \hat{m}}
\end{split}
\end{equation}
where $\hat{n}$ and $\hat{m}$ are the number operators acting on $a$ and $b$ registers, respectively. Thus, after the phase modulations, the Trojan horse evolves into
\begin{equation}
    \ket{\xi'_{xy}} = \hat{U}_x \otimes \hat{U}_y \otimes \mathbbm{1} \ket{\xi} =\sum_{n,m} c_{nm} e^{i(nx+my)\frac{\pi}{2}} \ket{n} \ket{m} \ket{\mathcal{E}_{nm}}
\end{equation}
which gives the effective state
\begin{equation}
    \ket{\phi_{xy}} = \ket{e^{ix \frac{\pi}{2}} \alpha} \ket{e^{iy \frac{\pi}{2}} \beta} \ket{\xi'_{xy}} 
\end{equation}
when Alice is given the input $x$ and Bob is given the input $y$. Hence, for a fixed combination $x,y,x',y'$, we have
\begin{equation}
\begin{split}
    \braket{\phi_{x'y'}|\phi_{xy}} = & \left(\sum_{nm} |c_{nm}|^2 e^{i(n(x-x')+m(y-y'))\frac{\pi}{2}} \right) \times \\&
    \braket{e^{ix' \frac{\pi}{2}}\alpha|e^{ix \frac{\pi}{2}}\alpha}\braket{e^{iy' \frac{\pi}{2}}\beta|e^{iy \frac{\pi}{2}}\beta}
\end{split}
\end{equation}
where the term in the parenthesis is the contribution due to the THA. Here, $\alpha$ and $\beta$ denote the amplitudes of Alice's and Bob's lasers.

Now, due to the symmetry of the phase modulations, observe that
\begin{equation}
    \begin{split}
        e^{i (n+4) x \pi/2} = e^{i n x \pi/2} e^{i2\pi x} = e^{inx\pi/2}\\
        e^{i (m+4) y \pi/2} = e^{i m y \pi/2} e^{i2\pi y} = e^{imy\pi/2}
    \end{split}
\end{equation}
for all $x,y \in \{0,1,2,3\}$. Hence, without loss of generality, it is sufficient to consider $n,m \in \{0,1,2,3\}$. Denoting the probability that Alice and Bob receive $n$ and $m$ photons respectively, $|c_{nm}|^2 = P_{nm}$, it is therefore sufficient to consider finite number of $\{P_{nm}\}_{n,m}$.

Here, in contrast to the analysis presented in Ref. \cite{primaatmaja2019versatile}, the Gram matrix of the effective signal states are not fixed as Eve can vary the photon number distribution to maximize the leakage. Therefore, there are two variables that we consider in our optimization, namely the Gram matrix of Eve's quantum side information (which we denote by $G$) and the photon number distribution (which we denote by $P_{nm}$). Therefore, taking into account Eve's freedom to choose the Trojan horse state, we have to solve the following optimization problem
\sloppy
\begin{widetext}
\begin{equation}
\begin{split}
\max_{G,P_{nm}} \quad & \eph\\
\text{s.t.} \quad & G \succeq 0,\\
& e_{\rm ph} \leq 1/2,\\
& P(z|x,y)= \braket{\mathrm{e}^z_{xy}|\mathrm{e}^z_{xy}},\\
& P_{nm}\geq 0,\\
& \sum_{n,m} P_{nm} =1, \\
& \sum_{n,m}{P_{nm} n}\leq \nu_A,\\
& \sum_{n,m}{P_{nm} m}\leq \nu_B, \\
& \sum_{z} \braket{\mathrm{e}^z_{x'y'}|\mathrm{e}^z_{xy}} = \sum_{nm} P_{nm} e^{i(n(x-x')+m(y-y'))\frac{\pi}{2}} \Lambda_{x'y',xy}
\end{split}
\end{equation}
\end{widetext}

where $n,m \in \{0,1,2,3\}$, $\nu_A$ and $\nu_B$ denote the intensity of the Trojan horse lights (measured at the output of Alice's and Bob's source, respectively) and
\begin{equation}
    \Lambda_{x'y',xy} = \braket{e^{ix' \frac{\pi}{2}}\alpha|e^{ix \frac{\pi}{2}}\alpha}\braket{e^{iy' \frac{\pi}{2}}\beta|e^{iy \frac{\pi}{2}}\beta}
\end{equation}
is the inner product of Alice's and Bob's characterized signal states (i.e. the signal states in the absence of THA).

One could then plug the bound on $\eph$ into the key rate formula
\begin{equation}
    R \geq P_\text{pass} [1 - h_2(\eph) - h_2 (e_\text{bit})]
\end{equation}
where $h_2(\cdot)$ is the binary entropy function and $e_\text{bit}$ is the bit error rate in the key generation basis.

\medskip

\nocite{*}


\end{document}